
\magnification=1200
\parskip 3 pt plus 1pt minus 1 pt
\parindent=0pt
\vsize=22 true cm
\hsize=15 true cm
\overfullrule=0pt
\def\p{\partial} \def\D{{\cal D}}  \def\bD{{\bar {\cal D}}}
\def\bw{{\bar w}}  \def\bW{{\bar W}}
\def\e{\epsilon} \def\g{\gamma} \def\n{\nabla} \def\d{\delta}
 \def\s{\sigma}  \def\z{\zeta} \def\be{{\bar \eta}}
\def\c{\chi} \def\t{\vartheta}  \def\bt{{\bar \vartheta}}
\def\b{\beta} \def\a{\alpha} \def\l{\lambda}  \def\f{\varphi}
\def\da{{\dot \alpha}} \def\db{{\dot \beta}}  \def\dg{{\dot \gamma}}
 
\def\sd{self-dual } \def\ssd{super self-dual }
\def\sdy{self-duality } \def\ssdy{super self-duality }
\def\sym{super Yang-Mills } \def\ym{Yang-Mills }
\def\rep{representation}
\def\half{{1\over 2}}
\def\der#1{{\partial \over \partial #1}}

\rightline{Bonn-HE-93-23}
\rightline{hep-th/9306163}
\vskip 1 true cm
\centerline{THE MATREOSHKA OF SUPERSYMMETRIC SELF-DUAL THEORIES}
\vskip 2 true cm
\centerline{Ch. Devchand$^*$  and  V. Ogievetsky$^*$
\footnote{}
{$^*$ on leave from the Joint Institute for Nuclear Research, Dubna,
Russia.}
}
\vskip .35 true cm
\centerline{{\it     Physikalisches Institut}}
\centerline{{\it       Universit\"at Bonn }}
\centerline{{\it       Nussallee 12}}
\centerline{{\it    D-53115 Bonn 1, Germany }}
\vskip 3 true cm
{\bf Abstract}

Extended super self-dual systems have a structure reminiscent of a
``matreoshka''. For instance, solutions for N=0 are embedded in solutions
for N=1, which are in turn embedded in solutions for N=2, and so on.
Consequences of this phenomenon are explored. In particular, we  present an
explicit construction of local solutions of the higher-N super self-duality
equations starting from any N=0 self-dual solution. Our construction uses
N=0 solution data to produce N=1 solution data, which in turn yields N=2
solution data, and so on; each stage introducing a dependence of the solution
on sufficiently many additional arbitrary functions to yield the most general
supersymmetric solution having the initial N=0 solution as the helicity +1
component. The problem of finding the general local solution of the $N>0$
super self-duality equations therefore reduces to finding the general
solution of the usual (N=0) self-duality equations. Another consequence of
the matreoshka phenomenon is the vanishing of many conserved currents,
including the supercurrents, for super self-dual systems.
\vskip 2 true cm
\vfill
\eject

{\bf 1. Introduction}
\vskip 15pt
Super Yang-Mills massless multiplets are well known to have components
classified by helicity (rather than spin) and consisting of two irreducible
Lorentz representations:

$$\matrix{   &helicity:&  \half   & 0   &  -\half   & -1 \cr
	     &  N=0   &          &     &           &  f_{\da\db} \cr
	     &  N=1   &          &     &  \l_\da   &  f_{\da\db} \cr
	     &  N=2   &          &  W  & \l_{\da i} &  f_{\da\db} \cr
	     &  N=3   & \c_\a    & W^i & \l_{\da i} &  f_{\da\db} \cr}
\eqno(1)$$ and

$$\matrix{   &helicity:&    1       &  \half   & 0 &  -\half    \cr
	     &  N=0   &  f_{\a\b}  &          &   &            \cr
	     &  N=1   &  f_{\a\b}  & \l_\a    &   &            \cr
	     &  N=2   &  f_{\a\b}  & \l^i_\a  & \bW  &           \cr
	     &  N=3   &  f_{\a\b}  & \l^i_\a  & W_i  & \c_\da  \cr}
\eqno(2)$$

where $f_{\a\b}, f_{\da\db}$ are the (1,0) and (0,1) components of the
field strength tensor,
$\l$'s and $\c$'s are spinor fields and the $W$'s are scalars.
In Minkowski space these two representations are related to each other
by CPT conjugation, however in spaces of signature (4,0), (2,2)
or complexified space, they are no longer conjugate to each
other, allowing one to set one of them to zero, the resulting equations
of motion being
precisely the \ssdy conditions (if the triangle of fields
in (1) is set to zero) or super anti-\sdy conditions (if the
the fields in (2) are set to zero). For instance, the equations
of motion for the full N=3 theory take the form [1]
$$\eqalign{
\e^{\da\dg} \D_{\a\dg} f_{\da\db} +& \e^{\b\g} \D_{\g\db} f_{\a\b}
=\ [\l_{\a i}, \l_\db^i ]  + [\c_{\a }, \c_\db ] + [ W_i ,\D_{\a\db} W^i ]
+ [W^i  ,\D_{\a\db} W_i ]\cr
& \e^{\dg\da} \D_{\a\dg} \l_{\da i} = - \e_{ijk} [\l_{\a}^j , W^k]
				     + [\c_{\a }, W_i ]   \cr
& \e^{\g\b} \D_{\g\db} \l_{\b}^i = - \e^{ijk} [\l_{\db j} , W_k]
				   + [\c_{\db }, W^i ] \cr
& \e^{\dg\da} \D_{\a\dg} \c_{\da } = - [\l_{\a}^k , W_k]  \cr
& \e^{\g\b} \D_{\g\db} \c_{\b} = - [\l_{\db k} , W^k] \cr
\D_{\a\db}\D^{\a\db} W_i & = - 2 [[W^j , W_i], W_j] + [[W^j , W_j], W_i]
  +\half \e_{ijk}\{\l^{\a j}, \l_\a^k\} + \{\l^\da_i , \c_\da \}   \cr
\D_{\a\db}\D^{\a\db} W^i & = - 2 [[W_j , W^i], W^j] + [[W_j , W^j], W^i]
  +\half \e^{ijk} \{\l^\da_j, \l_{\da k}\} + \{\l^{\a i} , \c_\a \}
\cr}\eqno(3)$$
Now setting the fields in (1) to zero yields the \ssdy equations
$$\eqalign{
\e^{\b\g} \D_{\g\db} f_{\a\b} =& 0\cr
 \e^{\g\b} \D_{\g\db} \l_{\b}^i =& 0 \cr
\e^{\dg\da} \D_{\a\dg} \c_{\da } =& - [\l_{\a}^k , W_k]  \cr
 \D_{\a\db}\D^{\a\db} W_i  =& \half \e_{ijk}\{\l^{\a j}, \l_\a^k\}
.\cr}\eqno(4)$$
The first equation is identically satisfied in virtue of the Bianchi
identity and the condition $ f_{\da\db} = 0$ which is just the usual
(N=0) first order \sdy equation for the vector potential:
$$ F_{\mu\nu} = \half\e_{\mu\nu\rho\s} F_{\rho\s}\ .\eqno(5)$$
The second equation is just a source-free Dirac equation for a triplet
of Weyl fermions which may be thought of as zero-modes of the covariant
Dirac operator in the background of a \sd vector potential. The third
and fourth equations are however {\it not} source-free and their
solutions do not allow interpretation as zero-modes of gauge-covariant
operators in the background of a \sd gauge potential. So the folklore
that all spinor and scalar fields in \sd supermultiplets are
merely such zero-modes, although true for the N=1 case,
is clearly false for higher N.
The corresponding equations for $N < 3$ follow on truncation of the
N=3 multiplet $\{ f_{\a\b}, \l^i_\a , W_i \equiv \e_{ijk}W^{jk}, \c_\da,
\c_\a , W^i \equiv \e^{ijk}W_{jk}, \l_{\da i} ,  f_{\da\db}\}$:
For N=2 the internal index i,j=1,2, $W^{ij} = \e^{ij}\bW ,
W_{ij}= \e_{ij} W, \c_\da=\c_\a=0 $; and for N=1 the internal index as
well as all $W$'s and $\c$'s drop out.

The full N=4 multiplet is an irreducible self-conjugate \rep. N=4 \sdy
therefore does not fit into the above scheme and the recently
discussed N=4 \ssd theory [2] is in fact not a restriction
of the full N=4 \sym theory, but an independent theory containing the
entire component multiplet of the latter.

The N=0 \ym \sdy conditions have recently drawn a great deal of renewed
interest, especially in view of the prospects of classifying lower
dimensional integrable systems in terms of their possible reductions
(e.g. [3]). Super \sdy equations (reviewed by e.g. [4])
open the remarkable new possibility
of obtaining further classes of lower dimensional integrable systems by
dimensional reduction than have hitherto been obtained from N=0 \sdy.
They have also been found to have a hitherto hidden
relation to N=2 superstrings, whose spectrum they reproduce [2,5].

The main purpose of this paper is to unravel the structure of local
solutions of  extended \ssd gauge theories in complexified superspace
([6,7]),
ignoring questions concerning appropriate reality conditions,
global properties and finiteness of the Yang-Mills action.
As we shall show, lower N solutions are embedded
in higher N ones in a manner reminiscent of a matreoshka. Clearly lower N
solutions may be obtained from higher N ones by truncation and solutions
of lower N self-duality conditions (including the non-supersymmetric
N=0 case) are manifestly solutions of the higher N \sdy conditions (with
extra component fields being zero). We ought to emphasize, however,
that not all higher N
solutions can be obtained by superconformal or super-Poincare transformation
of lower N solutions. We shall elucidate this embedding structure in
section 3, explicitly demonstrating how a higher N solution may be
iteratively constructed from any given
lower N solution. The general higher N solution thus constructed
depends on new arbitrary functions not present in the original solution.
It is therefore not merely a supersymmetry or even superconformal
transformation of the latter, since these transformations depend on
{\it parameters} rather than arbitrary functions.
Our method uses an iterative procedure for promoting lower N solutions to
higher N ones, thus reducing the main problem to that of solving the N=0
equations. Given any N=0 solution, its {\it general}
N=1 extension depending on an additional Lie algebra valued arbitrary
function can be constructed. Similarly, any N=1 solution can be promoted to
an N=2 solution depending on {\it two} extra Lie algebra valued arbitrary
functions; and any N=2 solution may be used to construct an N=3 solution
depending on three new arbitrary funtions. So up to N=3, {\it the number
of arbitrary functions doubles at each stage, just like the number of
degrees of freedom in the theory}. We thus obtain the most general
supersymmetric extension of any given solution of (5), making the solution
of for instance the last three equations in (4) a mere technicality once
(5) has been solved. In section 4 we shall
illustrate our procedure by constructing some particular examples of
N=1 extensions of the N=0 one-instanton solution.

Self-dual restrictions of
non-self-dual theories are remarkable not only for their solubility
but also in virtue of being theories in which the conserved currents of
the latter identically vanish in the former. Super self-duality,
just as ordinary (N=0) \sdy (5), implies the {\it source-free} second-order
Yang-Mills equations for the gauge vector field, i.e. it implies
the vanishing of the Yang-Mills source current.
We shall demonstrate the latter explicitly (for  $1 \le N \le 3$) in
section 4  by showing that the spin 1 source current
factorises into pieces which vanish for respectively \sd and anti-\sd
theories.
We shall also demonstrate that the supercurrents generating supersymmetry
transformations in pure supergauge theories,
also factorise into
parts which vanish for respectively \sd and anti-\sd solutions.
This is the generalisation to supersymmetric pure gauge theories of the
factorisation of the Yang-Mills stress tensor into \sd and
anti-\sd parts of the field-strength tensor. So just as \sdy implies the
vanishing of the pure Yang-Mills stress tensor, as a further
manifestation of the matreoshka phenomenon, \ssdy
implies the vanishing of the pure supergauge theory supercurrent
which contains the stress tensor and its superpartners.
Although we presently restrict our attention to super \ym theories,
the situation in self-dual supergravity appears to be very similar.
The \sdy condition for the Riemann tensor implies the source-free
Einstein equation, which is tantamount to the vanishing of the
supercurrents for the superpartners of the spin 2 field (the
non-gravitational sources).
These interrelationships are part of an intricate matreoshka of
self-dual gauge and supergravity theories which it is our purpose to
begin to unravel in this paper.

\vskip 20pt
{\bf 2. Self-dual super gauge potentials}
\vskip 15pt
We shall use the notation of [6] in which we introduced
harmonic superspaces with coordinates
$$\{ x^{\pm \a} \equiv u^\pm_\db x^{\a\db},\
  \bt^\pm_i \equiv u^\pm_\da \bt^{\da}_i ,\ \t^{\a i} ,\ u^\pm_\da \}
,$$ where $u^\pm_\da $ are harmonics (see e.g.[8,9]), and gauge covariant
derivatives defined in a {\it central basis} by
$$\eqalign{ \D_{\a i} &= D_{\a i} + A_{\a i} \cr
	 \bD^{+i} &= \bar D^{+i} + \bar A^{+i} = u^{+\da} \bD^i_{\da} \cr
   \n^+_\a &= \p^+_\a  + A^+_\a =  u^{+\da} \D_{\a\da }\  ,\cr}$$
where the u-dependence (linearity for $\n^+_\a, \bD^{+i}$ and independence
for $\D_{\a i}$) is better expressed in the form
$$ [ D^{++} , \D_{\a i} ]  = [ D^{++} , \bD^{+i}] = [ D^{++} , \n^+_\a ]
=  0\  ,\eqno(6)$$
where $D^{++} = u^+_\da \der{u^-_\da} $ is a harmonic space derivative.

The \ssdy conditions (i.e. the vanishing of the representations in (1))
are equivalent
to the consistency conditions for the following system of
linear equations in harmonic superspace
$$\eqalignno{ \D_{\a i} \f & = 0 &(7a) \cr
	     \bD^{+i} \f & = 0  &(7b)\cr
	     \n^+_\a \f &  = 0 ,&(8)\cr}   $$
i.e. the conditions [10]
$$\eqalign{
     \{\D_{\a i}, \D_{\b j}  \} = & 0 =  \{ \bD^{+i} , \bD^{+j} \}  \cr
	[\n^+_\a , \n^+_\b ]  =  & 0 =  [ \bD^{+i} , \n^+_\a ] \cr
\{ \D_{\a j} , \bD^{+i} \}& =  2 \d^i_j \n^+_\a  \cr
\{ \D_{\a i} , \n^+_\b  \}  & = 0, \cr} \eqno(9)$$

where the matrix function $\f$, which we take to have unit determinant,
takes values in the gauge group.
These harmonic spaces have analytic subspaces suitable
for describing \sd restrictions.
In fact, as we showed in [6], choosing an {\it analytic basis}
in which the gauge covariant
derivatives $\{ \D_{\a i}, \bD^{+i} , \n^+_\a  \}$ in (3-9) lose their
connections and $D^{++}$ instead acquires a connection component  defined by
$$V^{++} =  \f^{-1} D^{++} \f ,\eqno(10)$$
the \ssdy conditions may be reconstructed from solutions of
generalised Cauchy- Riemann conditions for a chiral analytic (i.e.
independent of the coordinates  $\{x^{-\a}, \bt_i^- , \t^{\a i} \}$)
superfield $V^{++}$:
$$\eqalign{ \der{\t^{\a i}}  V^{++} &= 0 \cr
 \der{ \bt_i^{-} }  V^{++} &= 0 \cr
  \der{ x^{-\a} }  V^{++} &= 0 .\cr}\eqno(11) $$
Eqs. (11) are just eqs. (6) in the {\it analytic basis} [6].
Given a $\f$ solving (10,11),
the components of the gauge potential
$$A^+_\a    =\   -\  \partial^+_\alpha \f \f^{-1}, \quad  \quad
  \bar A^{+i} =\    -\  \bar D^{+i}\f \f^{-1}
\eqno(12)     $$
automatically solve (9),
with (6) implying linearity of these potentials
in the $u's$, so the superspace gauge potentials $ A_{\a\db}, A^i_\da $
may be read off from the harmonic expansions
$$A^+_\a  =  u^{+\da} A_{\a\da} ,\quad \quad
\bar A^{+i} = u^{+\da} A^i_{\da}.\eqno(13)$$

There exists, therefore, a correspondence between analytic $V^{++}$'s and
\sd gauge potentials.
This is basically the supertwistor [11]
correspondence for \sd gauge theories [12].
In the next section we shall use this correspondence in order to construct
an explicit general form of the supersymmetric solution
in terms of any given solution of $N=0$ \sdy.
The crucial step is the construction of a $\f$ satisfying (10,11), which
transforms under the gauge group as [9]
$$ \f \rightarrow
e^{-\tau(x^{\a\da}, \bt_i^\da \t^{\a i})} \f
			e^{\l(x^{+\a}, \bt_i^+ , u^\pm_\da)}
\eqno(14)$$
so that the analytic $\l$ (we shall call any function independent of the
coordinates  $\{x^{-\a}, \bt_i^- \}$ `analytic') is the gauge
transformation parameter for the harmonic space potential $V^{++}$,
while the $\tau$ is the usual gauge transformation parameter for
the super gauge potentials $ A_{\a\db}, A^i_\da $. The matrix function
$\f$ is therefore a `bridge' [9] between these two realisations of the
gauge group.

For our considerations the following remark is crucial:

{\it The gauge freedom (14) and the constraints (9) allow us to choose
a chiral basis in which the bridge $\f$
in (7,8) is  independent of the grassmann variables $\t^{\a i}$ and
$\bt^-_i$, allowing us to replace the system (7,8) by the purely even system}
$$ \n^+_\a \f = 0 .\eqno(8)$$
The odd equations (7), in this basis, either being implied by (8)
or implying only explicitly resolved kinematic constraints.
This system (8) is identical to the linear system for standard (N=0)
\sdy ; the only difference is that now the bridge $\f$ is a superfield
depending on the grassmann variables $\bt^+_i$.
To prove the above statement we first note that in view of (9) we may
always choose a chiral basis (with connection components $A_{\a i} = 0$)
in which the derivatives in (4,9) take the form
$$\eqalign{ \D_{\a i} =&\ \der{\t^{\a i}} ,\cr
     \bD^{+i} =&\ \der{\bt^-_i}  + 2 \t^{\a i} \der {x^{-\a}} + A^{+i}\  ,\cr
	    \n^+_\a  =&\ \der {x^{-\a}} + A^+_\a   \  .\cr}\eqno(15)$$
Now the last two equations in (9) yield
$$\der{\t^{\a i}}\bar A^{+j} = 2 \d^i_j A^+_\a \eqno(16)$$
and $$\der{\t^{\a i}} A^+_\a = 0 ,$$
which imply a relationship between the vector and spinor potentials,
namely,
$$\bar A^{+i} = 2\  \t^{\a i} A^+_\a \ .$$
Now this relation together with (7,8)  in the basis (15) immediately yields
$$ \der{\bt^-_i}  \f = 0  ,$$
proving that the only odd dependence of $\f$ (and $A^+_\a$)
is that on $\bt^+_i$. (This was independently known to [13] too).
So (7) and (8) are actually not independent equations. In fact they
are equivalent (a rather peculiar property of the \ssd systems), since the
chirality of $\f$ means that (7) implies (8); and conversely,
the relationship (16) and the chirality
of $\f$ immediately  give (7) as a consequence of (8).
This equivalence is manifest in this chiral basis, in which we shall
henceforth work.

Now, in view of the above remarks, the superfield
$A_\a^+ = A_\a^+(x^\pm_\a, \bt^+_i)
= u^{+\da}A_{\a\da}(x_{\b\db},\bt_{\da i}) $
satisfying (9) contains all
the component fields in (2) satisfying the appropriate \ssd restrictions
of the equations of motion. For instance, the N=3 superfield vector
potential has expansion
$$A_{\a\db}(x, \bt) = A_{\a\db}(x) + \bt_{\db i} \l^i_\a(x)
		+ \e^{ijk} \bt_{\da j} \bt^\da_i \n_{\a\db} W_k(x)
      + \e^{ijk} \bt_{\da i} \bt^\da_j \bt^\dg_k \n_{\a\dg}\c_\db \ ,
      \eqno(17)$$
and the condition that $A_\a^+(x^\pm_\a, \bt^+_i)$ satisfies (9) is
equivalent to the component fields $\{ A_{\a\db}, \l^i_\a, W_k \}$
satisfying the \ssd restrictions (4) of the equations of motion (3).

\vskip 20pt
{\bf 3. The solution matreoshka}
\vskip 10pt
As we have seen, the problem of constructing \ssd potentials
reduces to finding a matrix function $\f$ of unit determinant satisfying
$$ D^{++} \f(x^{\pm\a}, \bt_i^+ , u^\pm_\da )
=  \f(x^{\pm\a}, \bt_i^+ ,u^\pm_\da ) V^{++}(x^{+\a}, \bt_i^+,u^\pm_\da )
,\eqno(18)$$
for an arbitrary analytic superfield $V^{++} $
taking values in the gauge algebra, from which the potentials can be
constructed using eqs.(12,13). In virtue of (18) $\f$ is a functional of
$V^{++} $ and on going from the theory with N-1 supersymmetries to the one
with N supersymmetries, the number of arbitrary functions in $V^{++} $
(i.e. the number of components in its $\bt$-expansion) doubles for $N \le 3$.
That a \sd solution for any given N also satisfies \sdy for any higher N
is manifest in terms of the analyticity conditions (11).
Lower N analyticity is clearly contained in higher N analyticity;
explicitly,
$$ V^{++}(x^{+\a}, \bt^{+i} ,u^\pm_\da)
= v^{++}(x^{+\a}, u^\pm_\da) +  \bt^{i+} v_i^+(x^{+\a},u^\pm_\da )
+ \bt^{i+}\bt^{j+}v_{ij}(x^{+\a},u^\pm_\da) + ..... , \eqno(19)$$
so an analytic function for some given N is manifestly an analytic function
for any higher N (with null higher components); and conversely,
from any N-analytic function, simple truncation yields
lower-N analytic functions.
Thus, not only is lower N analyticity (\sdy) embedded in higher N
analyticity (\sdy), but the higher N \sdy implies lower N \sdy for
appropriately truncated multiplets.
As we have seen, the general solution  may be expressed in terms of
the matrix  function $\f $, which in turn is a functional (in virtue of
eq.(18)) of the set of arbitrary analytic functions
$ \{ v^{++} , v^+_i , v_{ij} , .... \} $
taking values in the gauge algebra.
Remarkably, for any $N \le 3$ the number of these arbitrary analytic
functions is precisely equal to the number of degrees of freedom
contained in the fields in (2).
Now consider for instance N=0, for which
case the general solution depends on $v^{++}(x^{+\a})$. Supersymmetry
and superconformal transformations can indeed be used to promote
this to a solution of higher N \ssdy. However such transformations can only
be used to add a number of parameters to the solution; not dependences on
further arbitrary functions.
So the general (local) solution for
higher N may not be obtained by merely performing a superconformal
or supersymmetry transformation on the general solution for lower N.

Our main result is the construction of the solution of (18)
in terms of the components of $V^{++}$ in (19)  and
the general solution $ \f_b$ of the $N=0$ equation
$$ D^{++} \f_b(x^{\pm\a}, u^\pm_\da)
=  \f_b(x^{\pm\a},u^\pm_\da) v^{++}(x^{+\a},u^\pm_\da)  .\eqno(20)$$

We start with the N=1 theory for which we write the required
`bridge' $\f$ in the form
$$\f = e^{\bt^+ \psi^-(x^{\pm\a}, u^\pm_\da)} \f_b(x^{\pm\a},u^\pm_\da)
     = ( 1 + \bt^+ \psi^-(x^{\pm\a},u^\pm_\da)) \f_b(x^{\pm\a},u^\pm_\da)
,\eqno(21)$$
where $\f_b$ is some (presumed to be known) solution of (20).
Inserting this factorisation as well as the expansion
$$V^{++}(x^{+\a}, \bt^{+},u^\pm_\da)
= v^{++}(x^{+\a},u^\pm_\da) +  \bt^{+}v^{+}(x^{+\a},u^\pm_\da)$$
in (18), we obtain a first order equation for  $\psi^-$ :
$$ D^{++} \psi^- = \f_b v^+ \f_b^{-1} .\eqno(22)$$
For particular cases where the right hand side is a simple polynomial
in the x's and u's, this equation allows direct integration using the
differentiation rules
$$ D^{++} u^+_\da = 0  ,\quad D^{++}u^-_\da = u^+_\da $$
$$ D^{++} x^+_\da = 0  ,\quad D^{++}x^-_\da = x^+_\da .$$
The general solution to (22) may however be expressed in terms
the harmonic space Green function [14] solution to the equation
$$ D^{++} F^{(q)}  = G^{(q+2)} ,$$ namely,
$$ F^{(q)} = \int d^2u' {(u^+ u'^-)^{q+1} \over (u^+ u'^+)} G^{(q+2)}(u')
.\eqno(23)$$
Thus:
$$  \psi^-(x^{\pm\a}, u^\pm)
    = \int ~d^2u'  {1 \over (u_i^+ u'^{+i})} \f_b v^+ \f_b^{-1}
,\eqno(24)$$
determining an N=1 bridge $\f$ in terms of any given N=0 $\f_b$
and an arbitrary
analytic function $v_{1}^{+}(x^{+\a}, u^\pm)$. From this $\f$ the potentials
may  be constructed as indicated in the previous section, for instance
inserting (21) in (12) we obtain the following expresssion for the vector
potential:
$$\eqalign{ A^+_\a  =& -\ \partial^+_\alpha \f_b \f_b^{-1}
   -\  \bt^+ (\p^+_\a \psi^- - [\p^+_\alpha \f_b \f_b^{-1} , \psi^- ]) \cr
=& -\ \partial^+_\alpha \f_b \f_b^{-1} -\  \bt^+ \n^+_\a \psi^- \
.\cr}\eqno(25)$$
Eq.(3b) guarantees this to be linear in $u^+$,
so the \sd vector potential can be easily recovered.
As we explained in the introduction, the coefficient of $\bt^+$
in the above superfield vector potential is precisely the spinor
fields $\l_\a$ satisfying the Dirac equation in the background of
component vector potential
$ A^+_\a  = -\ \partial^+_\alpha \f_b \f_b^{-1} $.
We shall explicitly carry out this construction starting from some
particular examples of analytic $V^{++}$'s in the next section.

We now further dress this N=1 bridge (21) to an N=2 bridge, thus
constructing the next layer of the solution matreoshka.
Inserting the matreoshka-like ansatz for N=2
$$\eqalign{ \f &= e^{\bt^{2+} \psi^-_2(\bt^{1+}, x^{\pm \a})}
		 e^{\bt^{1+} \psi^-_1(x^{\pm \a})} \f_b  \cr
&= ( 1 + \bt^{2+} \psi_2^-(x^{\pm\a})
	+ \bt^{2+} \bt^{1+} \psi_{21}^{--}(x^{\pm\a}))
( 1 + \bt^{1+} \psi_1^-(x^{\pm\a})) \f_b(x^{\pm\a}) ,\cr}\eqno(26)$$
together with the expansion
$$V^{++}(x^{+\a}, \bt^{+i})
= \left( v^{++}(x^{+\a}) +  \bt^{+1}v_{1}^{+}(x^{+\a})  \right)
  + \bt^{2+} \left( v_{2}^{+}(x^{+\a}) + \bt^{1+}v_{21}(x^{+\a})\right) $$
in (18) yields the following system of equations, where we denote a
conjugation by $\f_b$ by a tilde, $\tilde f \equiv  \f_b f \f_b^{-1} $;
$$\eqalign{  D^{++} \psi_1^- &= \tilde v_1^+  \cr
	     D^{++} \psi_2^- &= \tilde v_2^+   \cr
D^{++} (\psi_{21}^{--} + \psi_1^- \psi_2^-)
&=   \psi_2^- \tilde v_1^+  - \psi_1^- \tilde v_2^+ + \tilde  v_{21}
 , \cr}$$
Functions $\psi_1^-$ and $\psi_2^-$ solving the first two equations are
given by integrals of the form (24) and again using the Green function
formula (23) the third equation has solution
$$ \psi_{21}^{--} =\  -\ \psi_1^- \psi_2^-
    + \int ~d^2u'  {1 \over (u^+ u'^+)(u^+ u'^-)}
\left(  \psi_{[2}^- \tilde v_{1]}^+ + \tilde v_{12}  \right).$$
The N=2 bridge, depending on two further arbitrary analytic functions
$v_2$ and $v_{12}$, is therefore completely determined, affording immediate
reconstruction of the \ssd component multiplet from
$$ A^+_\a  = -\ \p^+_\alpha \f_b \f_b^{-1} -\  \bt^+_i \n^+_\a \psi^{i-}
  -\  \bt^+_2 \bt^+_1 (\n^+_\a \psi_{21}^{--}
+ \{ \psi^-_2 , \n^+_\a \psi_{1}^- \} ).$$

The matreoshka may now be enclothed in the next layer.
For the $N=3$ case, from which the above $N=1,2$ cases follow by
truncation, we take the $\f$ to have the following matreoshka structure
$$ \f = e^{\bt^{3+} \psi^-_3(\bt^{2+}, \bt^{1+}, x^{\pm \a})}
	 e^{\bt^{2+} \psi^-_2(\bt^{1+}, x^{\pm \a})}
	 e^{\bt^{1+} \psi^-_1(x^{\pm \a})} \f_b , $$
where $\psi^-_3$ and $\psi^-_2$ are N=2 and N=1 superfields respectively;
this form clearly breaking the internal SU(3) invariance, just as
the N=2 ansatz (26) breaks the internal SU(2) invariance.
Expanding the superfield $\psi^-_2$ as in (26) and $\psi^-_3$ as follows:
$$\psi^-_3(\bt^+_2, \bt^+_1, x^{\pm \a})
= \psi^-_3(x^{\pm \a}) + \bt^{2+} \psi^{--}_{32}(x^{\pm \a})
		       + \bt^{1+} \psi^{--}_{31}(x^{\pm \a})
		       + \bt^{2+} \bt^{1+} \psi_{321}^{---}(x^{\pm \a}) ,$$
and the N=3 analytic superfield thus:
$$\eqalign{ V^{++}(x^{+\a}, \bt^{+i})
=& \left(   v^{++}(x^{+\a}) +  \bt^{+1}v_{1}^{+}(x^{+\a})
+ \bt^{2+} \left( v_{2}^{+}(x^{+\a}) + \bt^{1+}v_{21}(x^{+\a})\right)
\right) \cr
+& \bt^{3+} \left( v_{3}^{+}(x^{+\a}) + \bt^{1+}v_{31}(x^{+\a})
+ \bt^{2+} \left( v_{32}(x^{+\a}) + \bt^{1+}v_{321}^-(x^{+\a}) \right)
\right)  ,\cr}$$
again yields a system of equations for the unknown functions in $\f$,
namely,
$$\eqalign{   D^{++} \psi_i^- &= \tilde v_i^+  ,\quad i=1,2,3, \cr
D^{++} (\psi_{ji}^{--} + \psi_j^- \psi_i^-)
&=   \psi_{[i}^- \tilde v_{j]}^+   +\tilde v_{ji}
,\quad i,j=1,2,3; j > i  \cr
D^{++}(\psi_{321}^{---} + \psi_{32}^{--} \psi_1^- - \psi_{31}^{--} \psi_2^-
		 & + \psi^-_3 (\psi_{21}^{--} + \psi_2^- \psi_1^-) )\cr
= \tilde v_{321}^-   + \psi_1^- \tilde v_{32}
 & +  \psi_3^- \tilde v_{21} - \psi_2^- \tilde v_{31}
  + (\psi_{21}^{--} + \psi_2^- \psi_1^-) \tilde v_3^+ \cr
 & + (\psi_{32}^{--} + \psi_3^- \psi_2^-) \tilde v_1^+
   - (\psi_{31}^{--} + \psi_3^- \psi_1^-) \tilde v_2^+
\cr}$$
which are again integrable using the harmonic space Green function
formula (23), thus determining the unknown functions in the bridge $\f$
in terms of which the solutions of (4) afford immediately extraction from
$$\eqalign{
 A^+_\a  =&  -\ \p^+_\alpha \f_b \f_b^{-1} -\  \bt^+_i \n^+_\a \psi^{i-}
  -\ \sum_{j>i} \bt^+_j \bt^+_i (\n^+_\a \psi_{ji}^{--}
+ \{ \n^+_\a \psi_{i}^- , \psi^-_j \} ) \cr
& +   \bt^{3+} \bt^{2+} \bt^{1+} ( \n^+_\a \psi_{321}^{---}
+ [\n^+_\a \psi_{1}^{-} , \psi_{32}^{--}]
+ [\n^+_\a \psi_{2}^{-} , \psi_{31}^{--}]
+ [\n^+_\a \psi_{21}^{--} , \psi_{3}^{-}] \cr & \quad\quad\quad\quad
- [ \{ \n^+_\a \psi_{1}^{-} , \psi_{2}^{-} \}, \psi_3^- ] ). \cr}$$

\vskip 20pt
{\bf 4. Examples}
\vskip 10pt
In this section we illustrate our construction by generating some
particular N=1 solutions from some simple monomial examples of
the analytic spinorial function
$v^+$, starting from the well-known N=0 SU(2) BPST instanton solution
$$ A_{\a\da i}^j = {1\over \rho^2 + x^2}
		   (\half x_{\a\da} \d^j_i + \e_{i\a} x^j_\da),$$
which may easily be seen [15] to correspond to the analytic function
$$ (v^{++j})_i^j = {x^{+j} x^+_i \over \rho^2} \eqno(27)$$
via the bridge
$$ (\f_b)_i^j = \left( 1 + {x^2 \over \rho^2} \right)^{-\half}
		 \left( \d^i_j + {x^{+i}x^-_j \over \rho^2} \right) $$
having inverse
$$ (\f_b^{-1})_i^j = \left( 1 + {x^2 \over \rho^2} \right)^{-\half}
     \left[ \left( 1 + {x^2 \over \rho^2} \right)\d^i_j
	   - {x^{+i}x^-_j \over \rho^2}                  \right] ,$$
where i,j are SU(2) indices.
Using this N=0 data as the core of the solution matreoshka, the
procedure described in the previous section affords construction
of higher N solutions.
To illustrate the method, we now explicitly carry out this construction
for N=1 for four particular examples of analytic spinorial function $v^+$.
Since our gauge group is SU(2), these have to be chosen to be traceless.
We shall also use the {\it normal gauge} [8] which fixes part of the
$\l$ gauge freedom in (14) such that the analytic data does not have any
explicit $u^+$ dependence.
In this gauge (see [8] for a discussion)
an analytic function cannot be expressed as a total $D^{++}$ derivative
of another analytic function. It is therefore a convenient gauge which
avoids gauge artifacts by distinguishing gauge inequivalent $V^{++}$'s.
The above $v^{++}$ is clearly compatible with this gauge; and we shall choose
our $v^+$'s in it too.
\vskip 20pt
\item{a)}
We start with the simplest case of a $v^+$ linear in $x^+$: and having a
constant spinorial parameter $\z_i$ of dimension $[cm]^{-{3\over 2}}$
$$ (v^+)^j_i =  x^{+j}\z_i + x^+_i\z^j   \ . $$
This gives
$$\f_b v^+ \f_b^{-1} = \left( 1 + {x^2 \over \rho^2} \right)^{-1}
  \left( x^{+j} \z_i + \left( 1 + {x^2 \over \rho^2} \right) x^+_i \z^j
	 - { 1 \over \rho^2} x^+_ix^{+j}x^-_l \z^l
	 - { 1 \over \rho^2} x^-_ix^{+j}x^+_l \z^l  \right)
,$$
which affords integration to
$$\psi^{-j}_i = \left( 1 + {x^2 \over \rho^2} \right)^{-1}
  \left( x^{-j} \z_i + \left( 1 + {x^2 \over \rho^2} \right) x^-_i \z^j
	 - { 1 \over \rho^2} x^-_ix^{+j}x^-_l \z^l  \right)
.$$
{}From (25,13) the vector potential  may now be found to be
$$ A_{\a\da i}^j = {1\over \rho^2 + x^2}
		   (\half x_{\a\da} \d^j_i + \e_{i\a} x^j_\da)
    + \bt_\da  { \rho^4 \over (\rho^2 + x^2)^2}
		\left( \e_{i\a} \z^j + \d^j_\a \z_i \right)
.$$
In fact this is not a genuinely new solution since it is related
to the N=0  one we started with by a supertranslation with parameter
$\rho^2\z^\a $:
$$ x^{+\a} \rightarrow x^{+\a} + \bt^+  \rho^2\z^\a .$$

\vskip 20pt
\item{b)}
A $v^+$ linear in $x^+$ , but not a supertranslation of the N=0
$v^{++}$ (27) is :
$$ (v^+)^j_i =  x^{+p}c_{pik}\e^{kj}  , $$
where $c_{pik} $ is a totally symmetric tensor parameter having,
like the parameter $\z$ of the previous example, dimension
$([cm]^{-{3\over 2}})$.
This yields
$$\eqalign{
\psi^{-j}_i =&\  x^{-p} c_{pik}\e^{kj}
  + {1\over \rho^2}c_{pkn}\e^{nj}(  x^{+p}x^{-k}x^-_i
				    - \half x^{-p}x^{-k}x^+_i ) \cr
 & - {1\over (\rho^2 + x^2)} \left[ {c_{pik} \over 2} x^{-p}x^{-k}x^{+j}
    + {c_{pkn}\over \rho^2} ( \half x^{+p}x^{-k}x^{-n} x^{+j}x^-_i
	      - {1\over 6} x^{-p}x^{-k}x^{-n} x^{+j}x^+_i ) \right]
\cr}$$
and
$$ A_{\a\da i}^j = {1\over \rho^2 + x^2}
		   (\half x_{\a\da} \d^j_i + \e_{i\a} x^j_\da)
     + \bt_\da  c_{\a in}\e^{nj} \left( 1 + {x^2 \over \rho^2} \right)
.$$
This simple solution, however, does not vanish asymptotically.
\vskip 20pt
\item{c)}
Now choosing a $v^+$ quadratic in $x^+$ with constant spinorial
parameter $\be^\da$ of dimension $[cm]^{-{5\over 2}}$:
$$ (v^+)^j_i =  x^{+j} x^+_i u^-_\da \be^\da   , $$
we obtain
$$\psi^{-j}_i = \left( 1 + {x^2 \over \rho^2} \right)^{-1}
 (x^{+j} x^-_i u^-_\da + x^{-j} x^+_i u^-_\da - x^{-j}x^-_iu^+_\da)\be^\da
,$$ which yields the \sd vector potential
$$ A_{\a\da i}^j = {1\over \rho^2 + x^2}
		   (\half x_{\a\da} \d^j_i + \e_{i\a} x^j_\da)
	    + \bt_\da { \rho^4 \over (\rho^2 + x^2)^2}
		     ( \e_{i\a} x^j_\db  - \d^j_\a x_{i\db} )\be^\db
.$$
This solution is related to the N=0 one by asuperconformal transformation
with parameter $\rho^2 \be_\da$
$$ x^{+\a} \rightarrow  x^{+\a} ( 1 - \rho^2 \be_\da u^{-\da} \bt^+)$$
and is precisely the solution discussed by [16].
\vskip 20pt
\item{d)}
Considering a $v^+$ cubic in $x^+$, with constant spin-tensor parameter
$d_{pkl}$ of dimension $[cm]^{-{7\over 2}}$:
$$ (v^+)^j_i =  x^{+j} x^+_i x^{+p}u^{-k}u^{-l} d_{pkl}   $$
we find
$$\eqalign{
\psi^{-j}_i = \left( 1 + {x^2 \over \rho^2} \right)^{-1} d_{pkl}
	  \{   &  x^+_i x^{+j} x^{-p} u^{-k} u^{-l}
	     + x^-_i x^{-j} x^{-p} u^{+k} u^{+l}
	     - x^-_i x^{-j} x^{+p} u^{+k} u^{-l}     \cr
       & + (x^+_i x^{-j} + x^-_i x^{+j})
	      (x^{+p} u^{-k} u^{-l} - x^{-p} u^{+k} u^{-l}) \}
,\cr}$$
which yields
$$ \eqalign{ A_{\a\da i}^j =&\  {1\over \rho^2 + x^2}
		   (\half x_{\a\da} \d^j_i + \e_{i\a} x^j_\da) \cr
&\  +\  \bt_\da \  d_{pkl} \left[  \d^p_\a x^k_ix^{jl}
	 +  {\rho^2 \over \rho^2 +  x^2 }
\left( \e_{i\a} x^{jl}x^{pk}   + \d^j_\a x^l_ix^{pk} \right)  \right]
\cr}$$
This example, like (b), also does not vanish asymptotically, but is
yet another example of solution which is not related to the starting
N=0 solution by a supersymmetry or superconformal transformation.

Although we may similarly produce further solutions, or further promote
these to higher N solutions (as described in the previous section),
we restrict ourselves to these simplest examples of local solutions to
the N=1 \ssdy equations.
As explained before, the coefficients of $\bt$ in these N=1 superfield
vector potentials are precisely the spinor fields $\l_\a$ satisfying
the Dirac equation in the background of the N=0 instanton solution we
started with.

\vskip 20pt
{\bf 5. Vanishing supercurrents and source currents of \sd supermultiplets}
\vskip 15pt
As mentioned in the introduction, the usual (non-\sd) super \ym multiplets
(for $0 \le N \le 3$) consist of two irreducible
Lorentz representations; and
\ssdy corresponds to setting one of these to zero.
Many conserved currents of the full theory, for instance the spin 1
source currents as well as the supercurrents, factorise into
parts from the two representations. They therefore vanish for
\sd multiplets. This is a further matreoshkan phenomenon; the vanishing
of the currents being consequences of the simple fact that N=0 \sdy
automatically implies, on the one hand, the {\it source-free} \ym
equations, and on the other hand, the vanishing of the stress tensor.

\vskip 15pt
{\bf N=0}

In 2-spinor notation the components of the field-strength tensor
are defined by
$$[{\cal D}_{\alpha{\dot \alpha}} ~,~{\cal D}_{\beta{\dot \beta}}]
\equiv\ \epsilon_{\alpha\beta} f_{{\dot \alpha}{\dot \beta}}  ~
+~\epsilon_{{\dot \alpha}{\dot \beta}} f_{\alpha\beta}   $$
and the usual N=0 \sdy is simply the statement that the (0,1) component
$ f_{\da\db} $ vanishes,
leaving only the (1,0) field $ f_{\a\b} $, which satisfies
the source-free Yang-Mills equations $\e^{\g\a} \D_{\g\db} f_{\a\b}=\ 0 $
automatically in virtue of the Bianchi identity.
The stress tensor
clearly vanishes; the factorisation into \sd and anti-\sd fields
being manifest in spinor notation:
$$ T_{\alpha{\dot \alpha},\beta{\dot \beta}}
\equiv\ f_{{\dot \alpha}{\dot \beta}} f_{\alpha\beta} =\ 0.  $$

\vskip 15pt
\goodbreak
{\bf N=1}

The full $N=1$  on-shell Yang-Mills multiplet
$\{  f_{\da\db}, \l_\da, \l_\a, f_{\da\db}  \}$
satisfies the equations
$$\eqalign{
\e^{\dg\da} \D_{\a\dg} f_{\da\db} + \e^{\g\b} \D_{\g\db} f_{\a\b}
= [\l_\a, \l_\db]   \cr
\e^{\dg\da} \D_{\a\dg} \l_{\da} = 0 = \e^{\g\b} \D_{\g\db} \l_{\a} .\cr} $$
Now \ssdy (i.e. the vanishing of the fields $\{ f_{\da\db} , \l_\da  \}$
in the above equations)
manifestly implies that the vector source current
$$ j_{\a\da} \equiv [\l_\a, \l_\db] = 0 . $$

In terms of superfields the full
theory is conventionally described in terms of
two spinorial field strengths $w_\a, \bar w_\da $ having leading components
$\l_\a, \l_\da$ respectively; and defined by
$$\eqalign{
[\bD_{\da} ~,~\D_{\a\db} ] =& \e_{\da\db}  w_\a       \cr
[\D_{\b} ~,~\D_{\a\db}] =&\  \e_{\b\a} \bar w_\db  .\cr} $$
The supercurrent is given by the axial vector superfield [17]
$$ V_{\a\da} = w_\alpha \bar w_\da $$
satisfying the covariant conservation law
$$ \D^{\a \da} V_{\a\da} = 0$$
in virtue of the superfield equations of motion: covariant (anti-) chirality
conditions for $w_\a$ (resp. $\bar w_\da$).
Now the N=1 \sdy conditions take the form
$$\bar w_\db = 0, \eqno(28)$$
manifestly implying the vanishing of the supercurrent; and also implying
(via super Jacobi identities) the N=0 \sdy conditions for the spin 1 component.

\vskip 15pt
{\bf N=2}

The $N = 2$  Yang-Mills theory, on the other hand, may be described
in terms of two scalar superfields $W , \bar W$ defined by
$$\{\D_{\a i} ~,~\D_{\b j}\} =\e_{ij} \e_{\a\b} W        $$
$$  \{\bD^{i}_\da ~,~\bD^{j}_\db \} = \e^{ij} \e_{\da\db} \bar W  ,$$
in terms of which the supercurrent takes the form [18]
$$ V = W \bar W $$
and satisfies the conservation law
$$ \bD^{ij}V = 0 ;\quad \bD^{ij} = \bD^{i[+}\bD^{-]j} $$
in virtue of the equations of motion
$$ \bD^{ij} W = 0 = \D^{ij} \bW   $$
and Bianchi identity
$$ \D^{ij} W + \bD^{ij} \bW = 0 . $$
N=2 \sdy in superfield form
$$ W = 0 $$ clearly implies the former in virtue of the latter.
It manifestly implies that the supercurrent $V$ vanishes and
it also implies the vanishing of the spinor superfield
$\bw^{i}_\da \equiv \bD^i_\da W  $; so even the superfield form of
N = 1 \sdy (28) is embedded in N = 2 \sdy.

The component theory is described by the
equations of motion
$$\eqalign{
\e^{\dg\da} \D_{\a\dg} f_{\da\db} + \e^{\g\b} \D_{\g\db} f_{\a\b}
&= [\l_{\a i}, \l_\db^i ]  + [W ,\D_{\a\db} \bW ]
+ [\bW ,\D_{\a\db} W ]\cr
\e^{\dg\da} \D_{\a\dg} \l_{\da i} &= - [\l_{\a i} , W]    \cr
 \e^{\g\b} \D_{\g\db} \l_{\a i} &=- [\l_{\db i} , \bW] \cr
\D_{\a\db}\D^{\a\db} \bW  &= [\l^\a_i, \l_\a^i]  \cr
\D_{\a\db}\D^{\a\db} W &= [\l^\da_i, \l_\da^i] \cr} $$
(where we denote the leading components of the scalar superfields
$W, \bW$ by the same symbols).
Clearly, for N=2 \sd configurations  ( when the multiplet
$\{ f_{\da\db} , \l^i_\da , W \}$ vanishes), the spin 1 and
spin $\half$ source currents vanish:
$$\eqalign{
j_{\a\db}  &\ \equiv  [\l_{\a i}, \l_\db^i ]  + [W ,\D_{\a\db} \bW ]
+ [\bW ,\D_{\a\db} W ] = 0 \cr
j_\a^i &\ \equiv [\l_{\a}^i , W] = 0 \cr
j_{\da i} &\ \equiv [\l_{\da i} , \bW] = 0 .\cr}$$

\vskip 15pt
{\bf N=3}

{}From the N=3 component equations of motion quoted in the introduction
(3), we see that for \sd configurations the following spin 1 and
spin $\half$ source currents vanish:
$$\eqalign{
j_{\a\db}  &\ \equiv  [\l_{\a i}, \l_\db^i ]  + [\c_{\a }, \c_\db ]
+ [ W_i ,\D_{\a\db} W^i ] + [W^i  ,\D_{\a\db} W_i ] = 0 \cr
j_\a^i &\ \equiv - \e_{ijk} [\l_{\a}^j , W^k]
		   + [\c_{\a }, W_i ] = 0 \cr
j_{\da i} &\ \equiv - \e^{ijk} [\l_{\db j} , W_k]
				   + [\c_{\db }, W^i ] = 0 .\cr}$$

In superfield language, the $N =3$  Yang-Mills equations may be written
in terms of two Lorentz-scalar superfield triplets
$W^{i}\equiv \e^{ijk}W_{jk} ,~~ \bW_{i} \equiv \e_{ijk}\bW^{jk}$
(again identically denoting superfields and their leading components)
defined by
$$\{\D_{\a i} ~,~\D_{\b j}\} =\ \e_{ijk} \e_{\a\b} W^k    $$
$$  \{\bD_\da^{i} ~,~\bD_\db^{j}\} =\ \e_{\da\db} \e^{ijk} \bW_{k}  ,$$
with supercurrent [18]
$$ V^i_j = W^i \bW_j - {1\over 3} \d^i_j W^k\bW_k ,$$
conserved in virtue of the superfield equations of motion:
$$  \D_{\a i} W_{jk} = \e_{ijk} w_\a  ,\quad
    \D^i_\da \bW^{jk} = \e^{ijk} \bw_\da  ,$$
$$    \D_{\a i} \bW^{jk} = \d_i^{[j} \l^{k]}_\a  ,  \quad
     \D^i_\da W_{jk} = \d^i_{[j} \l_{k]\da} .$$
Again, the superfield \sdy conditions
$$ W^i = 0 $$ imply vanishing of the supercurrent.

%
%
%
\goodbreak
\vskip 20pt
{\bf 5. Remarks}
\vskip 15pt
\item{a)} In conclusion we should like to stress once again that the
our method of solution reveals all {\it local} solutions of the \ssdy
equations. To find solution which are globally defined on $S^4$
is a separate problem. For instance supertranslations
and superconformal transformations of N=0 instantons yield such
solutions [16]. However, in physical applications one often needs
solutions having some
other reasonable asymptotic behaviour than that consistent with
regularity on $S^4$; for instance, solutions with periodic boundary
conditions corresponding to solutions on a 4-torus rather than a
4-sphere [19] or multiply-valued solutions on $S^4$ (e.g. [20]) which
presumably correspond to analytic $V^{++}$'s on harmonic spaces with
cuts.
\item{b)} The \sd N=4 theory recently discussed [2] in relation
to superstrings, being also described by the constraints (9),
is also amenable to solution by a straightforward
enlargement of our solution matreoshka.
\item{c)} Our procedure may also be applied to obtain static
solutions of the \ssdy equations, i.e. to obtain solutions of
the superspace Bogomolny equations, in particular \ssd monopoles.
We shall present these solutions elsewhere.
\item{d)}An interesting question is how further reduction to
soluble two dimensional
systems such as the (super-) KdV hierarchies manifests itself in terms of
the harmonic space analyticity conditions. Recent results on the embedding
of two dimensional soliton systems [3] in the \sdy conditions yield the
tantalising prospect that our solution matreoshka continues in the sense
that solutions of the two dimensional systems correspond to just further
truncations of the analytic data.
\vskip 15pt
We are grateful to A. Galperin, E. Ivanov, H. Nicolai, O. Ogievetsky,
A. Schwarz and E. Sokatchev for useful
discussions and to the
Physikalisches Institut, Universit\"at Bonn for hospitality.
One of us (VO) gratefully acknowledges receipt of
a Humboldt Forschungspreis which enabled this work to be carried out
in Bonn.
\vskip 15pt
{\bf References}
\vskip 15pt
\item{[1]} J. Harnad, J. Hurtubise, M. Legar\'e, S. Shnider,
Nucl. Phys. B256 (1985) 609.
\item{[2]} W. Siegel, Phys. Rev.D 46 (1992) R3235.
\item{[3]} L. Mason and G. Sparling, J. Geom. Phys. 8 (1992) 243.
\item{[4]} D. Amati, K. Konishi, Y. Meurice, G. Rossi, G. Veneziano,
Phys. Rep. 162 (1988) 169.
\item{[5]} H. Ooguri and C. Vafa, Nucl. Phys. B361 (1991) 469.
\item{[6]} C. Devchand and V. Ogievetsky, Phys. Lett. B297 (1992) 93.
\item{[7]} C. Devchand and A. Leznov, hep-th/9301098, IHEP 92-170.
\item{[8]} A. Galperin, E. Ivanov, V. Ogievetsky, E. Sokatchev,
Ann. Phys. (N.Y.) 185 (1988) 1;
M. Evans, F. G\"ursey, V. Ogievetsky, Phys. Rev. D47 (1993) 3496.
\item{[9]} A. Galperin, E. Ivanov, S. Kalitzin, V. Ogievetsky, E. Sokatchev,
Class. Quant. Grav. 1 (1984) 469.
\item{[10]}
A. Semikhatov, Phys.Lett. 120B (1983) 171;
I. Volovich, Phys.Lett. 123B (1983) 329.
\item{[11]} A. Ferber, Nucl.Phys. B132 (1978) 55;
\item{[12]}  R.S. Ward, Phys. Lett. 61A (1977) 81.
\item{[13]} O. Ogievetsky, private communication.
\item{[14]} A. Galperin, E. Ivanov, V. Ogievetsky, E. Sokatchev, Class.
Quant. Grav. 2 (1985) 601.
\item{[15]} A. Galperin, E. Ivanov, V. Ogievetsky, E. Sokatchev, in {\it
Quantum Field Theory and Quantum Statistics}, vol.2, 233 (Adam Hilger,
Bristol, 1987); JINR preprint E2-85-363 (1985);
S. Kalitzin and E. Sokatchev, Class. Quant. Grav. 4 (1987) L173;
O. Ogievetsky, in {\it Group Theoretical Methods in Physics},
 Ed. H.-D. Doebner et al, Springer Lect. Notes in Physics 313 (1988) 548.
\item{[16]} V. Novikov, M. Shifman, A. Vainshtein, M. Voloshin,
V. Zakharov, Nucl. Phys. B229 (1983) 394.
\item{[17]} S. Ferrara and B. Zumino, Nucl. Phys. B87 (1975) 207;
V. Ogievetsky and E. Sokatchev, Nucl. Phys. B124 (1977) 309;
Yad. Fiz. 28 (1978) 825; W. Lang, Nucl. Phys. B150 (1979) 201.
\item{[18]} P. Howe, K. Stelle, P. Townsend, Nucl. Phys. B192 (1981) 332.
\item{[19]} G. 't Hooft, Comm. Math. Phys. 81 (1981) 267.
\item{[20]} A. Zhitnitsky, Nucl. Phys. B340 (1990) 56;
D. Korotkin, Comm. Math. Phys. 134 (1990) 397.
\goodbreak
\end